# Energy gaps in high-transition temperature cuprate superconductors

Makoto Hashimoto,[1] Inna M. Vishik,[2,3,4*] Rui-Hua He,[2,3,4†] Thomas P. Devereaux,[2,3] and

Zhi-Xun Shen[2,3,4]

[1]*Stanford Synchrotron Radiation Lightsource,*

*SLAC National Accelerator Laboratory, Menlo Park, California 94025, USA*

[2]*Stanford Institute for Materials and Energy Sciences,*

*SLAC National Accelerator Laboratory, Menlo Park, California 94025, USA*

[3]*Geballe Laboratory for Advanced Materials, Stanford University, Stanford, California*

*94305, USA*

[4]*Departments of Physics and Applied Physics, Stanford University, Stanford, California*

*94305, USA*

Correspondence should be addressed to M.H. and Z.X.S.

* Present Address: Department Physics, Massachusetts Institute of Technology, Cambridge, MA, 02139, USA

† Present address: Department of Physics, Boston College, Chestnut Hill, MA 02467, USA

**The spectral energy gap is an important signature that defines states of quantum matter:**

**insulators, density waves, and superconductors have very different gap structures. The**

**momentum resolved nature of angle-resolved photoemission spectroscopy (ARPES) makes it a powerful tool to characterize spectral gaps. ARPES has been instrumental in establishing the anisotropic *d*-wave structure of the superconducting gap in high-transition temperature ($T_c$) cuprates, which is different from the conventional isotropic *s*-wave superconducting gap. Shortly afterwards, ARPES demonstrated that an anomalous gap above $T_c$, often termed the pseudogap, follows a similar anisotropy. The nature of this poorly understood pseudogap and its relationship with superconductivity has since become the focal point of research in the field. To address this issue, the momentum, temperature, doping, and materials dependence of spectral gaps have been extensively examined with significantly improved instrumentation and carefully matched experiments in recent years. This article overviews the current understanding and unresolved issues of the basic phenomenology of gap hierarchy. We show how ARPES has been sensitive to phase transitions, has distinguished between orders having distinct broken electronic symmetries, and has uncovered rich momentum and temperature dependent fingerprints reflecting an intertwined & competing relationship between the ordered states and superconductivity that results in multiple phenomenologically-distinct ground states inside the superconducting dome. These results provide us with microscopic insights into the cuprate phase diagram.**

# 1. Introduction

## 1.1 Cuprates

High-transition temperature ($T_c$) superconductivity in copper oxides (cuprates) is one of the most intriguing emergent phenomena in strongly-correlated electron systems.[1] It has attracted great attention since its discovery by Bednorz and Muller in 1986[2] because $T_c$ can exceed the boiling temperature of liquid nitrogen[3]. This not only implies broader possibilities for applications, but also is much higher than the putative limit of $T_c \sim 30$ derived from Fermi-liquid-like metals described by the Bardeen–Cooper–Schrieffer (BCS) theory.[4]

The cuprates have a layered crystal structure consisting of $CuO_2$ plane(s) (Fig. 1a inset) separated by charge reservoir layers that control the carrier concentration in the $CuO_2$ plane(s). As shown in the phase diagram (Fig. 1a), upon doping holes, the antiferromagnetic Mott insulating (AFI in Fig. 1a) phase of the parent compounds disappears and superconductivity emerges. $T_c$ follows a dome-like shape as a function of doping with a maximum $T_c$ around 16% doped holes per $CuO_2$ plaquette (optimally doped). A similar phase diagram is seen when electrons are doped into the parent compound, albeit with a more robust antiferromagnetic phase and a lower $T_c$.[5] On the hole doped side, there exists an enigmatic normal state above $T_c$ called the pseudogap, which will be the main focus of this article. The cuprate phase diagram, and particularly its unusual normal state, is unconventional compared to superconductivity in Fermi-liquid-like metals.

The cuprates exhibit a quasi two-dimensional electronic structure due to the quasi two-dimensional crystal structure. Figs. 1b and 1c show the schematic band structure and Fermi surface for the cuprates, respectively. A single band, originating mainly from hybridized *Cu $3d_{x2-y2}$* and *O $2p_{x,y}$* orbitals, crosses the Fermi level ($E_F$) (Fig. 1b), typically forming a large hole-like Fermi surface (Fig. 1c). It has been shown by angle-resolved photoemission spectroscopy (ARPES) in 1993[6] that the superconducting gap opens on the Fermi surface with a strong momentum anisotropy (*d*-wave gap), which will be discussed later in this review.

One of the greatest difficulties in formulating a microscopic theory of cuprate superconductivity is that the normal state above $T_c$ across a large part of the phase diagram is not a Fermi liquid. The state above $T_c$ but below a characteristic temperature $T^*$ that can be determined from many experiments features a spectral gap of yet unknown nature and has been labeled as the "pseudogap state"[7] (Fig. 1a), although its identification as a true thermodynamic phase is still a subject of debate. $T^*$ becomes larger with decreasing hole concentration, which is the opposite doping dependence from $T_c$ in the underdoped region ($p < 0.16$). The pseudogap was discovered as the "spin gap",[8] which manifests itself as an anomaly at $T^*$ in the spin-lattice relaxation rate of nuclear magnetic resonance (NMR), suggesting the suppression of the density of states around $E_F$ below $T^*$. Afterwards, anomalies at $T_c < T < T^*$ have been found by various experimental techniques including transport, scattering, and spectroscopic measurements.[7] Particularly, the momentum structure of the pseudogap was revealed by ARPES in 1996[9-11]. As high $T_c$ superconductivity emerges from this pseudogap state as the temperature is lowered, the pseudogap has been suggested to be intimately connected to the mechanism of high-$T_c$ and should provide clues of how even a higher $T_c$'s can be achieved.

In this article, we show how ARPES has contributed to the understanding of the cuprates. We first introduce in Chapter 1 the superconducting gap and the pseudogap in the ARPES spectra. In Chapter 2, we show the systematic doping and temperature dependence of the gap functions in momentum space, and present its interpretation. In chapter 3, we pay special focus to recent results[12-14] in understanding the pseudogap due to some order (the pseudogap order) distinct from superconductivity, which may be consistent with various symmetry breakings in the pseudogap state observed by different experimental techniques. We show evidence for a phase transition into a pseudogap phase at $T^*$ having broken electronic symmetry that is distinct from superconductivity. Well below $T_c$, we discuss how the pseudogap order is intertwined & entangled with superconductivity, which suggests multiple phenomenologically distinct ground states with non-trivial phase boundaries within the superconducting dome. These results provide us with deeper insights into the cuprate phase diagram, renewing the foundation for further study towards the complete understanding of the high-$T_c$ mechanism.

## 1.2 Angle-resolved photoemission spectroscopy (ARPES)

ARPES has been a leading tool to study gap anisotropies discussed in this article because it directly measures the occupied part of the single-particle spectral function[15] with ever increasing energy and momentum resolution. The cuprates are well-suited for the ARPES technique because of their quasi-2D electronic structure. The $Bi_2Sr_2CaCu_2O_{8+\delta}$ (Bi2212) and $Bi_2Sr_{2-x}La_xCuO_{6+\delta}$ (Bi2201) families in particular feature pristine cleaved surfaces which protect the low-energy bulk electronic structure, due to the weak Van der Waals between the two Bi-O planes. Over the past two decades, experiments have improved tremendously (Figs. 2a, 2c and 2d), allowing more precise information about electronic structure, including the gap functions, to be obtained (Fig. 2e). One recent development is the use of narrow-bandwidth UV lasers as light sources for photoemission. [14,16-21]. The superior resolution of laser ARPES allows unprecedented access to the lowest energy excitations near the node as shown in Figs. 2d and 2e. Additionally, traditional synchrotron-based ARPES continues to be improved with brighter synchrotrons and more powerful spectrometers. Synchrotron-based experiments have the advantage of covering a larger region of momentum space with photon energy flexibility. When one combines modern synchrotron and laser-based

ARPES experiments, one can gain deep insights into the nature of energy gaps as reviewed in this article.

## 1.3 Superconducting gap

In conventional BCS superconductors, an energy gap $\Delta_{SC}$ opens below $T_c$ with s-wave symmetry and minimal momentum dependence. $2\Delta_{SC}$ is the energy required to break a Cooper pair of electrons which form the superconducting condensate. In contrast, the superconducting gap in the cuprates is characterized by a strong momentum dependence. Early debates came to the conclusion that the superconducting gap function is consistent with an order parameter having $d_{x2-y2}$ symmetry, with support from ARPES,[6] penetration depth[22], Raman,[23] and phase sensitive measurements.[24] The d-wave symmetry of the superconducting gap has become an accepted fact when one constructs theories and interprets experimental results.

On the Fermi surface (Figs. 1c), the gap is the largest at the antinode – Fermi momentum ($k_F$) on the Brillouin zone boundary near ($\pi$, 0) (Fermi angle $\theta$ = 0°). The gap size gradually decreases towards the node along the Fermi surface and becomes zero at the node – $k_F$ in the CuO bond diagonal direction ($\theta$ = 45°). The superconducting gap changes sign across the node (Fig. 2b). We note that the terminology, 'node' and 'antinode' is still used above $T_c$ to refer to those particular regions of the Fermi surface. An ARPES study on Bi2212 by Shen et al.[6] was one of the key experiments which clarified the superconducting order parameter in the cuprates. This study compared energy distribution curves (EDCs) at two characteristic momenta, the node and the antinode, above and below $T_c$ to show the anisotropy of the gap. At the antinode, the opening of a gap below $T_c$ was detected as a leading edge shift of the EDC to higher binding energy and the emergence of a sharp quasiparticle peak at $\Delta_{sc}$ ~ 30 meV (upper EDCs in Fig. 2a). In contrast, the spectrum at the node does not show a shift in the leading edge gap across $T_c$ and the sharpening of the spectrum at low temperature is predominantly of thermal origin (lower EDCs in Fig. 2a).

## 1.4 Pseudogap

The pseudogap at $T_c < T < T^*$ also shows a strong momentum anisotropy in the occupied states, and thus, ARPES is a powerful experimental tool for investigating this central enigma in the cuprates. In Fig 3, we show the systematic momentum, temperature, and doping dependence of the spectral gaps in Bi2212. As shown in Fig. 3a, at $T << T_c$ in an underdoped $T_c$ = 92 K Bi2212 sample (denoted as UD92), the ARPES spectra along the Fermi surface show a strong momentum dependence. The EDCs are symmetrized with respect to $E_F$ to visualize the existence of a gap relative to $E_F$. The symmetrization removes the Fermi-Dirac cutoff, under the assumption of particle-hole symmetry, which is valid for a superconducting gap at $k_F$ and convenient for this part of the discussion. Momenta where symmetrized EDCs show a single peak at $k_F$ are said to be ungapped, and gapped momenta are characterized by a dip at $E_F$ in the symmetrized EDCs. The magnitude of the gap in the superconducting state was extracted by fitting these EDCs to a phenomenological model of the spectral function.[25] The superconducting gap is plotted as a function of Fermi angle θ in blue symbols in Fig. 3c, and the gap follows a simple *d*-wave form (blue dashed curve).

We show in Fig. 3b the symmetrized EDCs along the Fermi surface at $T > T_c$. In an ideal *d*-wave superconductor, one expects the superconducting gap to retain its *d*-wave form and to close uniformly at all momenta at $\sim T_c$. In contrast to this expectation, only a portion of the EDCs around the node shows a single peak at $E_F$ at $T > T_c$ as a signature of a vanishing gap. The EDCs around the antinode still show a gap feature. The gap function above $T_c$ obtained by the same fitting procedure as that for at $T < T_c$ is overlaid in Fig. 3c (red dashed line). The pseudogap, an anomalous gap above $T_c$, shows a strong momentum anisotropy that has some similarity to the *d*-wave form of the superconducting gap, but with an extended ungapped region around the node, defined as the "Fermi arc", which will be discussed later.  In ARPES, the pseudogap temperature $T^*$ is defined as the temperature when the gap at the antinodal $k_F$ closes. In Fig. 3e, the temperature dependence of the spectra at the antinode of UD92 is shown as an example where $T^* \sim 190$ K. The existence of the antinodal gap above $T_c$ was a surprise as it is not a simple metallic normal state that forms the basis for BCS theory. As an energy gap has special meaning to a superconductor, the pseudogap is an obstacle which must be first understood.

## 2. Nodal-antinodal dichotomy of the gaps

### 2.1 Temperature dependence of the gaps

The key characteristics of the pseudogap and the relationship between the pseudogap and superconductivity have been extensively investigated by systematically studying the gap function in different dopings, temperatures, and materials. Because of the completeness of the data set, we mainly discuss in this section the gap function in Bi2212.

Figure 3 shows overview of typical energy gaps in Bi2212 as a function of momentum (first row), temperature (second row), and doping (third row). To highlight the nodal-antinodal dichotomy of the energy gap, the first column represents the energy gap in the near-nodal region while the second column represents energy gap in the antinodal region. The extracted gap magnitudes, determined by analyzing the symmetrized spectra, as functions of momentum, temperature and doping are plotted in the third column for a more quantitative comparison

Figure 3e shows a typical temperature dependence of the antinodal spectra. One can clearly see that the gap does not close at $T_c$.[14,26] A similar behavior has been consistently observed by different ARPES groups over a wide doping range in different materials.[10-14,25-62] In addition, it has been reported that the intensity of the antinodal quasiparticle peak for underdoped samples shows a marked temperature dependence analogous to the superfluid density and becomes harder to detect above ~$T_c$.[30-34] These results are very different from conventional superconductors where a well-defined quasi-particle peak for a Fermi-liquid normal state exists above $T_c$, and suggest that superconductivity exists in an unconventional manner at $T < T_c$ in the antinodal region regardless of the nature of the pseudogap.

On the other hand, the gap in the near-nodal region appears simpler. First, the symmetrized spectra show a gap closing at ~ $T_c$ (Fig. 3d).[26] Additionally, the gap is particle-hole symmetric as the upper quasiparticle peak is observable above $E_F$ in the raw spectra slightly below $T_c$ (inset of Fig. 3d),[26,63] which has the same spectral weight as the lower quasiparticle peak below $E_F$ when the effects of the Fermi cutoff are accounted for. Although there have recently been extensive debates about what temperature the near-nodal gap closes (see also Box 1),[19,62,64-66] these signatures in the temperature dependence near the node are reminiscent of the BCS theory, suggesting that it is primarily attributable to superconductivity.

In the intermediate region between the antinode and the node, the energy gap becomes smaller with temperature, but does not close completely above $T_c$ (Fig. 3f). This suggests that the dichotomy in momentum space may not be sharp, but it is crossover-like. Because the gap in the antinodal and intermediate regions do not fully close at $T_c$ in the symmetrized spectra, only a portion of the Fermi surface is recovered near the node above $T_c$. This ungapped portion has been identified as the "Fermi arc"[28] (Fig. 3c). The phenomenology of this Fermi arc was presented in the first ARPES paper that reported the existence of the pseudogap.[9]

The antinodal pseudogap above $T_c$ eventually closes at a higher temperature, $T^*$, as shown for Bi2212 UD92 in Fig. 3d. This suggests that the full Fermi surface is recovered $T > T^*$. The recovery of the full Fermi surface above $T^*$ has been also observed in optimally doped Bi2201[12,13] which has simpler band structure because it lacks bilayer splitting. Here, the electronic structure above $T^*$ is well described by a single band tight-binding model, similar to that of simple metals.[67] We show in Fig. 4a and 4b the spectra above $T^*$ and well below $T^*$, respectively, for Bi2201. The nodal-antinodal dichotomy discussed above (also seen in Fig. 4c) no longer exists above $T^*$.

## 2.2 Doping dependence of the gaps

The doping dependence of the gap is another important piece of information on the nature of the pseudogap, which has been of particular interest over the past decade. In Fig. 3g and 3h, we show the doping dependences of the ARPES spectra at two characteristic momenta in the ground state: near the node ($\theta \sim 38°$) and near the antinode ($\theta \sim 0$). In this dataset,[14,26,68] it is clear that the doping dependence has a strong contrast between the two momenta. While the antinodal gap is strongly doping dependent, following the doping dependence of $T^*$, the gap near the node is nearly doping independent in the underdoped to optimally-doped regime.

To evaluate this dichotomy in momentum space in detail, we show in Fig. 3i the magnitude of the gap at $T \ll T_c$ as a function of the $d$-wave form factor, $|\cos(k_x)-\cos(k_y)|/2$, for various doping levels, with data from different doping offset by fixed amounts for clarity. For the most overdoped sample OD65, the magnitude of the gap is zero at the node and varies linearly as a function of the

*d*-wave form factor, indicating that the gap has a simple *d*-wave form. In the heavily overdoped regime, the magnitude of the gap changes along with the doping dependence of $T_c$, and the entire gap function maintains a simple *d*-wave form. However, as the doping level decreases, the near-nodal gap is surprisingly independent of doping until the deeply underdoped region is reached[14]. As a result, the gap function near the node approximately follows a simple *d*-wave form over the entire doping range. However, the gap function in the antinodal region deviates from this momentum dependence in the underdoped region and the magnitude of the antinodal gap approximately scales with *T**. Similar deviations of the gap function from a simple *d*-wave form in the ground state have been consistently reported in other cuprate compounds[34,36,38-40,44-46,53-55,59,60,69-74] (see also Fig. 4c for Bi2201).

## 2.3 Pseudogap due to some order distinct from superconductivity

The anomalous momentum, temperature and doping dependence of the gap function shown in Fig. 3 have been often interpreted as evidence for the coexistence of two distinct orders (although there are other interpretations as this continues to be a matter of debate as discussed in Box 1). In this picture, the distinct gap phenomenologies between the near-nodal and antinodal regions are due to different momentum, temperature, and doping dependence of the order parameters for the pseudogap and superconductivity which may coexist in the superconducting state. [12-14,26,38,40,44,45,51,52,54,55,57,59,62,68,71-75] At $T \ll T_c$, the gap in the antinodal region is more strongly affected by the pseudogap order which is distinct from superconductivity, while the *d*-wave gap in the near-nodal region is dominantly determined by superconductivity. The strong nodal-antinodal dichotomy of the gap function in the temperature dependence (Fig. 3a-f) can be understood if we allow for the coexistence of two different order parameters that have different momentum structures and temperature dependences. The stronger deviation from a simple *d*-wave form observed in the deeply underdoped regime (Fig. 3i) is explained by the larger discrepancy between the magnitudes of the two order parameters, where the magnitude of the pseudogap order parameter overwhelms that of superconducting order parameter in the antinodal region. In the more overdoped regime, the deviation is not as evident at the lowest temperature (Fig. 3i) because the magnitude of the pseudogap order parameter is small, and the fact that competition from superconductivity has weaken the pseudogap[14,76] (see Section 3.4). Nevertheless, the gap function deviates from a simple *d*-wave form at higher temperature (Fig. 3f), suggesting that the antinodal gap at higher temperature is more strongly affected by the distinct pseudogap order. One of the reasons these subtle effects could be observed is due to the improved resolution of synchrotron-based ARPES experiments (routinely <10 meV now vs. 20 meV in the past). Poorer resolution makes it difficult to accurately assess gaps especially when they are small. Here we note that the deviation of the gap function from a simple *d*-wave form in Bi2212 presented in Fig. 3 is unlikely to be an artifact due to stronger background signal in the antinodal region because quasiparticle peaks clearly exist over the entire Fermi surface even in UD50.[75] Furthermore, the doping in UD75 has been achieved only by annealing the as-grown Bi2212 without cation substitution, which might have some effect on the antinodal gap (also see Box 1).[14,51,75] At $T > T_c$, the contribution of superconductivity to the spectra is not strong and the gap function is explained by the dominating pseudogap order.

Consistent with these results from ARPES measurements, increasing results from various other experimental techniques have suggested a distinct electronic symmetry from that of superconductivity in the pseudogap state.[12,13,39,77-100] In particular, recent scattering experiments have clearly shown that charge modulations that compete with superconductivity universally exist in the bulk in the various hole-doped cuprate families.[101-111] Although the role of superconducting fluctuations above $T_c$ remains another important question in the field as discussed in Box1, the existence of some order associated with the pseudogap, which is distinct from superconductivity, is very likely an intrinsic and universal aspect of the high-$T_c$ cuprates, which may be important for understanding high-$T_c$ mechanism. In the next chapter, we will focus on the recent progress in this aspect, as it has been one of the most important subjects in the high-$T_c$ field in recent years. Electronic states having distinct signatures of the distinct pseudogap order are carefully examined by ARPES in expanded scopes beyond the energy gap on the Fermi surface.

## 3. Phase competition between the pseudogap and superconductivity

### 3.1 The pseudogap phase with a distinct electronic symmetry

In this chapter we show recent ARPES studies on Bi2201 and Bi2212 that suggest that the pseudogap order possesses a distinct electronic symmetry and competes with superconductivity. We first show the temperature dependence of the ARPES spectra at the antinode across $T^*$ in an optimally-doped Bi2201 sample. Here Bi2201 was chosen to study the pseudogap physics because of the wide temperature range for the pseudogap state (~100 K difference between $T^*$ and $T_c$). Also, mode-coupling effects near the antinode are weaker and the absence of the bilayer splitting[112] makes the data interpretation simpler. In optimally-doped Bi2201[12,13] ($T_c$ ~ 35 K), the peak position of the spectra at the antinodal $k_F$ is at $E_F$ above $T^*$, but continuously shifts below $E_F$ upon lowering temperature with a clear onset at $T^*$ (Fig. 5a). Here the raw ARPES spectra are divided by the Fermi-Dirac function convolved with resolution to effectively remove the effects of the Fermi cutoff and recover the single-particle spectral function in the vicinity of $E_F$. It has been also shown that the temperature dependence of the spectral weight at the antinode has a clear onset at $T^*$[12].

These abrupt changes at $T^*$ imply a phase transition to a broken-symmetry state below $T^*$. In fact, polar Kerr effect (PKE) measurements, sensitive to symmetry breakings, and time-resolved reflectivity (TRR) measurements on the same Bi2201 crystals show identical onset temperatures for the pseudogap phase transition (Fig. 5b).[13] The TRR signal for the pseudogap order also has the opposite sign to the one for superconductivity,[13] pointing to the different natures of the pseudogap and superconductivity. The analogous temperature dependences have been observed by both bulk and surface sensitive techniques and both the measurements in equilibrium and non-equilibrium states, supporting the idea that $T^*$ is associated with a phase transition to a non-superconducting broken-symmetry state (the pseudogap ordered state). In particular, consistencies between the previous PKE study[86] on $YBa_2Cu_3O_{7-\delta}$ (YBCO) and TRR studies[113-117] on YBCO and Bi2212 suggest that the phase transition at $T^*$ may be a general characteristic phenomena in the cuprates.

In Fig. 5c, the antinodal dispersions along the Brillouin zone boundary for $T > T^*$ and $T \ll T^*$ in nearly optimally-doped Bi2201 are compared.[12,13] The parabolic dispersion at $T > T^*$ clearly crosses $E_F$ and no gap exists, which allows us to define $k_F$ without ambiguity as indicated by red dashed lines. At $T \ll T^*$, the entire dispersion is pushed down to higher binding energy due to the opening of the pseudogap. There is no anomaly at $k_F$ in the dispersion in the pseudogap phase. Instead, the dispersion shows back-bending momenta markedly away from $k_F$ as indicated by green arrows. When a superconducting gap opens at $T < T_c$ (Fig.5d), one expects $k_F$ and the back-bending momenta (indicated by green arrows, noted as $k_G$) of the dispersion to be aligned. This is because homogeneous superconductivity requires a gap to open with particle-hole symmetry. This disconnection between $k_F$ and $k_G$ leads to the conclusion that the observed dispersion below $T^*$ arises from a different ordered phase than homogeneous superconductivity[12]. The smooth temperature evolution of the spectra at the antinodal $k_F$ (Fig. 5a) with onset-like behavior at $T^*$ suggests that the pseudogap phase transition at $T^*$ is responsible for the observed dispersion for the pseudogap order at $T \ll T^*$ as a signature of broken particle-hole symmetry.

It has been shown that the simple mean-field simulations for different types of density wave orders with bond-diagonal (Fig. 5e) and bond-directional wave vectors can qualitatively reproduce the misalignment between the back-bending momentum and $k_F$ as well as the strong downward shift of the band bottom energy that hardly happens for superconductivity (Fig. 5d).[12] Additionally, introducing a short correlation length of the density wave order can account for the anomalous broadening of the spectra (Figs. 4a and 5e) [12,13] with decreasing temperature that is opposite from the temperature dependence of the near-nodal spectra[13] (Fig. 4b) and opposite from expectation of thermal broadening. More detailed momentum-dependence study is required to clarify their relationship. The density wave orders considered in the simulation may be consistent with other experimental observations[39,79,82,89,93,95-99,101-111] and theoretical predictions[118] that suggest broken translational symmetry. Particularly, recent observations of bulk charge density wave in various cuprate compounds[101-111] suggest that broken translational symmetry is a general feature in the cuprates. Resonant inelastic x-ray scattering (RIXS) and x-ray diffraction support the idea of charge density wave (CDW) ordering tendency, with a fluctuating energy scale less than 100 meV.[101-111] Note that, the other side of the pocket[43,50,58] with broken particle-hole symmetry around the node,[43,58] have been claimed by some ARPES studies, but its relevance to the competing pseudogap order in the present discussion has not been clear yet.

## 3.2. Interplay between the pseudogap order and superconductivity

With a better understanding of the distinct pseudogap phase, one can gain deeper insights into the states below $T_c$ where the pseudogap and superconductivity coexist and interact. The ARPES spectra for the antinodal cut along the Brillouin zone boundary at $T \ll T_c$ for optimally doped Bi2201[13] are shown in Fig. 4a. The spectra are characterized by multiple energy features, whereas an ordinary homogeneous superconducting gap, which has a single energy feature below $E_F$, opens in the near-nodal region below $T_c$ (similar to near nodal behavior in Fig. 4b). The dispersion for the antinodal cut at $T \ll T_c$ obtained from Fig. 4a is summarized in Fig. 4d. The back bending of the dispersion at higher binding energy (blue open circles) dominantly due to the pseudogap order is misaligned to $k_F$, suggesting that the pseudogap exists in the superconducting state. The flat dispersion near $E_F$ at $T \ll T_c$ (green filled circles) quickly loses its definition above $T_c$, indicating that the flat dispersion is primarily of superconducting origin, similar to the quasiparticle peak in the antinodal region of Bi2212.[13] Also, fluctuating superconductivity, although playing a possible role in the vicinity of $T_c$,[119-125] does not appear to be responsible for the pseudogap phase that onsets at $T^* \gg T_c$. These two features in Fig. 5d cannot be explained only by homogeneous superconductivity or by two gaps simply adding in quadrature. Rather, the spectral function develops a complex structure with two energy scales of mixed origin in the ground state.

A simple mean-field model simulation for the coexistence of $d$-wave superconductivity and a density wave order[13] captures the observed spectral features in the optimally-doped Bi2201. As shown in Fig. 4e, this simulation captures: 1) the multiple energy features suggesting the coexistence of the order parameters for the pseudogap and superconductivity, 2) the misalignment between the back-bending momentum ($k_G$) and $k_F$ as a signature of the non-superconducting broken symmetry nature of the pseudogap order, 3) the anomalously flat low-energy dispersion dominantly assigned to superconductivity, and 4) the deviation of the gap function from a simple *d*-wave form.[13] The simulation suggests that the magnitudes of the order parameters for the pseudogap and superconductivity have comparable energy scales. Note that the order parameters cannot be directly determined from the observed energy scales (the shoulder and hump features). This is because these energy features are affected both by the pseudogap order and superconductivity in an unknown manner. Therefore, further theoretical studies are required to determine the magnitudes of the order parameters. Moreover, the flatness of the dispersion observed both in Bi2201 and Bi2212[13] suggests that superconductivity is distorted by the pseudogap order, which might partially contribute to the disappearance of the quasiparticle interference in the antinodal region reported by STS studies[53].

## 3.3 Distinct ground states in the superconducting dome

The intimate interplay of the order parameters for the pseudogap and superconductivity raises a question about how the character of the ground state changes with doping. By utilizing laser-based ARPES setup with superior momentum and energy resolution, it has been suggested that the superconducting dome consists of two or more phenomenologically distinct ground states.[14] Detailed gap measurements in the near-nodal region of Bi2212 allow for a more precise determination of the gap slope $v_\Delta$ (see Fig.6a for sample data), which measures how fast the *d*-wave gap increases as a function of momentum away from the node.

The doping dependence of $v_\Delta$ has been determined over a broad doping range at $T \ll T_c$. In Fig. 6e, $v_\Delta$ near the node is plotted as a function of doping. Surprisingly, $v_\Delta$ of ~39 meV appears to be independent of doping over a wide doping region below $p \sim 0.19$ (blue symbols) regardless of the large $T_c$ differences (40 - 96 K) as one can see in Fig. 6a that the gap functions near the node overlap well between different doping levels.[14] This suggests that $v_\Delta$ is not the only factor determining $T_c$, and the effect of the pseudogap order also needs to be taken into account. The doping-independent $v_\Delta$ over a wide doping range has not been fully resolved, but is supported by specific heat measurements in YBCO[126] and by STS data in Bi-based cuprates,[127,128] which could be an important key to further understand the interplay between the pseudogap order and superconductivity. In stark contrast, at $p > 0.19$, both $v_\Delta$ and $T_c$ decrease together as normally expected (Fig. 6e), suggesting that the pseudogap order is absent in ground state at $p > 0.19$, and superconductivity exists alone. Moreover, the change of $v_\Delta$ at $p \sim 0.19$ is abrupt, suggesting the sudden disappearance of the pseudogap order at $p \sim 0.19$.

This indicates a phase boundary at $T = 0$ as illustrated in Fig. 6d. The existence of this phase boundary has been suggested by both superfluid density measurements and the Cu-site impurity-doping needed to suppress superconductivity, both of which reach maxima at $p = 0.19$ [126,129,130]. These observations are robust in different materials, and suggest that a phase boundary at $p \sim 0.19$ could be a universal phenomenon across the cuprate families. Therefore, it has been concluded that there are at least two distinct ground states inside the superconducting dome. One is where superconductivity and the pseudogap order coexist ($p < 0.19$) and the other is where d-wave superconductivity exists alone ($p > 0.19$).

In addition, distinct physics is present in the deeply underdoped regime, with the most interesting observation being that the Fermi surface is gapped at every momentum.[131-133] This may indicate a third distinct ground state at $p < 0.076$[14] (Shaded region A in Fig. 6e). Interestingly, other cuprates also exhibit distinct phenomenology on the underdoped edge of the superconducting dome. Neutron scattering[134] and transport experiments[135,136] in YBCO suggest a critical doping 8-10%. Additionally, ARPES experiments on Bi2201[137], LSCO,[131,138,139] and Na-CCOC[131] also show a gap at the nodal momentum in underdoped samples, with the latter two systems showing this behavior at superconducting dopings. Various proposals have suggested the existence of such a critical point, including a metal-insulator quantum critical point,[135] a Lifshitz transition,[136] spin density wave order (SDW)[134], topological superconductivity,[140] Fulde–Ferrell–Larkin–Ovchinnikov (FFLO),[141] or coulomb gap.[142] Further studies are needed to ascertain if separate quantum critical points exist at $p \sim 0.19$ and $0.076$.[14]

## 3.4 Phase competition and the revised phase diagram

There have been conflicting reports in the literature about the endpoint of $T^*$ in the phase diagram. Some experiments suggest that it ends inside the superconducting dome, which is supported by the observation of the $T = 0$ phase boundary of the pseudogap order at $p = 0.19$.[60,126,129,130] In contrast, some spectroscopies, including ARPES, have reported the existence of the pseudogap order above $T_c$ even at $p > 0.19$.[143-146] Fig. 6f shows the antinodal spectra slightly above $T_c$ for different doping levels. In this finite temperature data, it is clear that the pseudogap above $T_c$ at the antinode exists at $p > 0.19$, different from that the influence of the pseudogap order on $v_\Delta$ at $T \ll T_c$ disappears at $p \sim 0.19$ as discussed in the previous section. This means that, for $0.19 < p < 0.22$, the pseudogap can exist at a finite temperature even though the ground state at $T = 0$ may be explained purely by superconductivity. Whether the antinodal gap completely closes at $T > T_c$ at $p > 0.22$ needs to be further examined with improved experimental setups.

These seemingly contradictory observations were reconciled by ARPES experiments showing that the pseudogap order is suppressed by superconductivity at low temperatures. This implies a phase diagram as schematically shown in Fig. 6d, and described in detail in Ref. [14]. $T^*$ decreases with doping until it touches the superconducting dome at $p > 0.22$ at $T = T_c$. Then, the pseudogap phase boundary "bends back" and goes into the superconducting, reaching $T = 0$ at $p \sim 0.19$ in the ground state. This proposed re-entrant phase diagram is naturally expected when considering the competition between two orders[91,147,148] as such a phase diagram has been reported in an iron-based high-$T_c$ compound $Ba(Fe_{1-x}Co_x)_2As$.[149,150] In Figs. 6a-6c, the gap functions of three dopings (UD40, UD65, and UD92) are compared at well below $T_c$, slightly below $T_c$, and slightly above $T_c$. At 10K (Fig.6a), doping-independent gaps extends to the intermediate momenta (boxed region) from near the node, while gaps around the antinode increase with underdoping. Just below $T_c$, however, doping-dependent gaps extend into the intermediate momenta. Slightly above $T_c$, gaps increase with underdoping everywhere except the Fermi arc near the node. As doping-independent (doping-dependent) gaps are a characteristic feature for superconductivity (the competing pseudogap order), the result in the intermediate momenta clearly suggests that the pseudogap order below $T_c$ is non-static and suppressed by superconductivity in a momentum dependent manner, supporting the proposed phase diagram with the re-entrant behavior of the pseudogap phase boundary in the superconducting dome.[14] Further, a more recent ARPES study[76] on the antinodal spectral weight has found that the spectral weight shows a singular behavior at $T_c$, which reveals an opposite sign of the impacts from pseudogap and superconductivity on the spectral weight. This is conclusive spectroscopic evidence for the competition between the order parameters for the pseudogap and superconductivity. The singular behavior of the spectral weight exists up to at least $p \sim 0.22$ in Bi2212, supporting that the pseudogap phase boundary bends back in the superconducting dome. Such a picture is supported by the recent finding of the suppression of the competing order below $T_c$ in x-ray scattering,[105,107,108,110] including Bi2212.[151] Further, the momentum region consistent with superconductivity-dominated physics gets larger with increasing doping.[14] This suggests that there is a doping regime where superconductivity might suppress the pseudogap order not just in a portion of the Fermi surface, but completely. This appears to happen $0.19 < p < 0.22$ in Bi2212,[14] resulting in the proposed phase diagram in Fig. 6d.

## 4. Summary and Outlook

This article has addressed the nature of the energy gaps using ARPES. In Chapter 2, we overviewed the strong nodal-antinodal dichotomy of the gap function, after introducing the pseudogap and superconducting gap in Chapter 1. In Chapter 3, we showed some recent ARPES studies that suggest the existence of a distinct pseudogap phase accompanied by broken electronic symmetry; and reveal the microscopic picture of the interplay between the pseudogap order and superconductivity. These results suggest that it is imperative to understand the competition of multiple phases and their universality in order to solve the high-$T_c$ mechanism. Scrutinizing and understanding the proposed phase diagram in Fig. 6e, together with further progress in the understanding of superconducting fluctuations above $T_c$, will be an important step forward. For a comprehensive understanding of the pseudogap physics, further investigations in real and reciprocal spaces as well as results from phase sensitive probes are warranted. Such efforts could provide us with deeper insights into various novel phenomena in complex oxides, as the competition among various broken-symmetry states is a recurring theme of strongly-correlated physics.

## Box1: Fluctuating superconductivity above $T_c$

There is general consensus in the field that the gap in the antinodal region does not close at $T_c$. It may be opened by pairing, competing orders, or coexistence of them, as the existence of energy gap alone does not tell us its nature. This has created varying interpretations regarding the nature of the pseudogap observed as the antinodal gap by ARPES. The existence of the competing pseudogap order has become very likely due to recent experimental progresses discussed in this article. However, as another possibility for the origin of the antinodal pseudogap, fluctuating superconductivity or preformed incoherent Cooper pairs above $T_c$[11,28,30,32,35,37,41-43,47-49,55,56,58,61,69,70,152,153] has been an important focus of the field, which could be consistent with other experimental results[119-125,154] such as the Nernst effect measurements.[121] In this case, $T^*$ represents the temperature where Cooper pairs begin to form, and $T_c$ represents the temperature where phase coherence is achieved. The nodal-antinodal dichotomy in the temperature dependence (Fig. 3a-f) is understood by the strong momentum dependence of this preformed Cooper pair forming above $T_c$. This picture is supported by the reports that show that the antinodal gap evolves smoothly across $T_c$, suggesting that the nature of the antinodal gaps above and below $T_c$ are closely related (Fig. 3d and 3f). Further, it has been suggested that the antinodal and near-antinodal gaps are particle-hole symmetric[42,43,58,63,155], which is required for the opening of homogeneous superconducting gap thus consistent with the one gap picture. A superconducting gap which maintains a simple *d*-wave form down to the lowest dopings is another argument for this picture. There have been reports indicating precisely that in Bi2212 in the wide doping[61] down to a deeply underdoped $T_c$ = 0 K sample[153] and in other cuprates.[39,41,47-49,55,74,152] Particularly, it has been recently argued that the cation substitution affects the size of the antinodal gap, suggesting that the deviation of the gap function from a simple *d*-wave from in the antinodal region may not be a universal and intrinsic phenomenon by the pseudogap, but it may be material specific.[61] This argument is contradicted by other observation that shows the same deviation from non-cation substituted samples[156]. Experimental conditions[55,74,156], cation substitutions[56,61,156], out-of-plane disorder[40,54,71], and sample quality are all considerations which must be more systematically evaluated in converging on this portion of the debate.

The momentum-space structure of the pseudogap—particularly the Fermi arc— is another crucial information to address the nature of the energy gaps. An unclosed Fermi arc is highly nontrivial because a Fermi surface cannot terminate within the BZ zone. Some experiments were interpreted as that the Fermi arc corresponds to a thermally broadened nodal gap structure.[28,35,37,41,49,55,157] This was first examined by studying the temperature dependence of the Fermi arc length in the pseudogap and scaling the results for all dopings with their respective $T^*$. It was found that this scaled arc length extrapolates to zero in the limit of zero temperature.[35,49] These results imply a node in the ground state of the pseudogap, which is more consistent with the fluctuation above $T_c$. Notably, some recent results more directly suggest that a *d*-wave gap above $T > T_c$ around the node.[62,64-66] The detailed temperature and momentum dependence studies suggest that there is a characteristic temperature $T_{pair}$ ($T_c < T_{pair} < T^*$) where a *d*-wave gap opens.[62] The integrated ARPES spectra perpendicular to the Fermi surface suggest that the energy gap is insensitive to $T_c$ and the scattering rate increases rapidly across $T_c$ around the node.[64-66] These results derived from varying materials, methodology and extrapolation, should be investigated further to fully understand the Fermi arc phenomenology. .

It is also interesting to point out that there have been several reports on the coexistence of signatures of simple *d*-wave superconducting gap and a competing order at higher energy in the antinodal region,[39,55,74] which may reconcile some aspects of the contradictory pictures regarding the pseudogap nature. Particularly, it has been reported that both the gaps for a competing order and superconductivity persist above $T_c$.[55,57,62] Whether the two orders coexist microscopically[12,13] or not, how they are entangled if they coexist, how the ARPES background signal contributes to the spectra around the antinode,[73] and what temperature the fluctuation goes away[19,57,62,64-66], are important open questions. Taking the evidence in totality, the existence of a competing order is an important feature of the pseudogap physics, while the extension to which fluctuating superconductivity contributing to the pseudogap physics is an open issue that should be investigated further.

## Additional Information

Correspondence and requests for materials should be addressed to M.H. and Z.X.S.

## Acknowledgement

We acknowledge Y. L. Chen, K. Tanaka, W.-S. Lee, and B. Moritz for sharing their ARPES data and for making figures; and A. Fujimori, Z. Hussain and D.H. Lu for long term collaboration. This work is supported by the Department of Energy, Office of Basic Energy Science, Division of Materials Science.

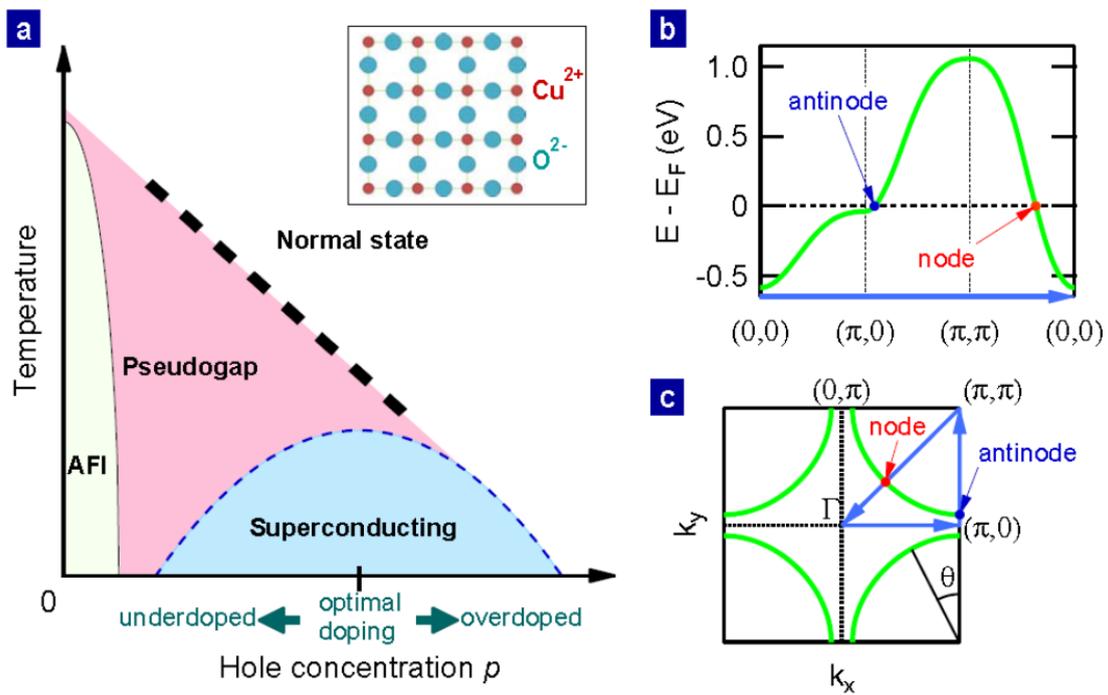

**Fig. 1 High-$T_c$ cuprate superconductors: a** Schematic phase diagram. Inset shows the crystal structure of the $CuO_2$ planes, which are of central relevance to superconductivity and the pseudogap. **b** Schematic band dispersion in reciprocal space for cuprates along the high-symmetry cuts as shown in blue in **c**. **c** Fermi surface, where the nodal and antinodal momenta and the Fermi angle θ are defined.

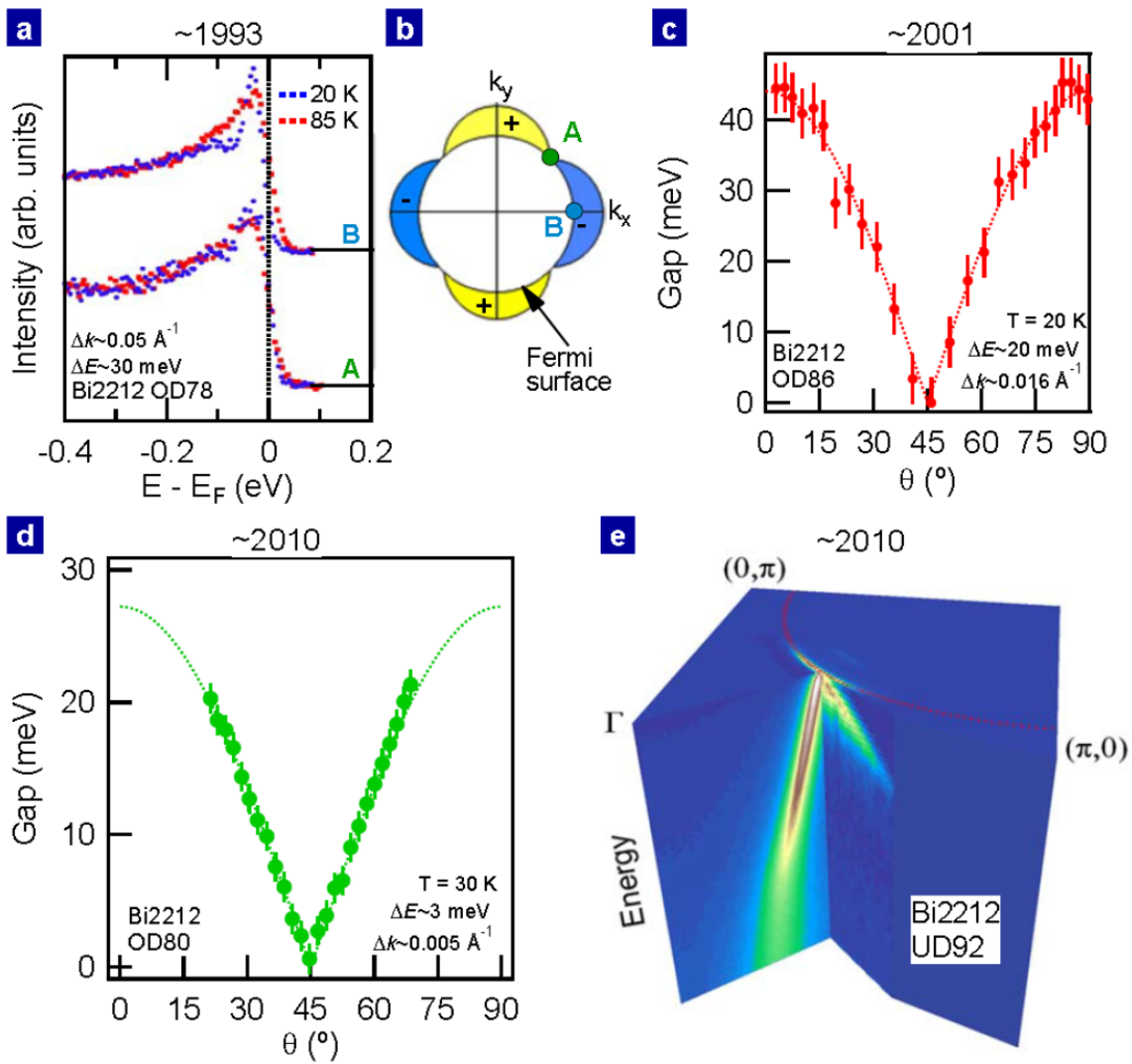

**Fig. 2 *d*-wave superconducting gap symmetry in cuprates witnessed by ARPES and the improvement of ARPES data quality: a** Superconducting gap anisotropy first observed in 1993 (reproduced from Ref. [6]). Lower spectra (A) are taken at the node and upper spectra (B) at the antinode. **b** Schematic of a *d*-wave order parameter on a circular Fermi surface. Gap is zero at the node where the superconducting gap changes sign (A) and maximum at the antinode (B). **c** Typical synchrotron gap measurement a decade ago as a function of the Fermi angle θ. Error bars in uncertainty of determining $E_F$ (±0.5 meV), error from fitting procedure, and an additional 100% margin. **d** Near-nodal gaps measured by a modern laser-based ARPES system with superior resolutions and high photon flux. Error bars reflect 3σ error in fitting procedure and an additional 100% margin. **e** Three dimensional ARPES data set, showing the quasiparticle dispersions both perpendicular and parallel to the Fermi surface near the node, reproduced from Ref. [16].

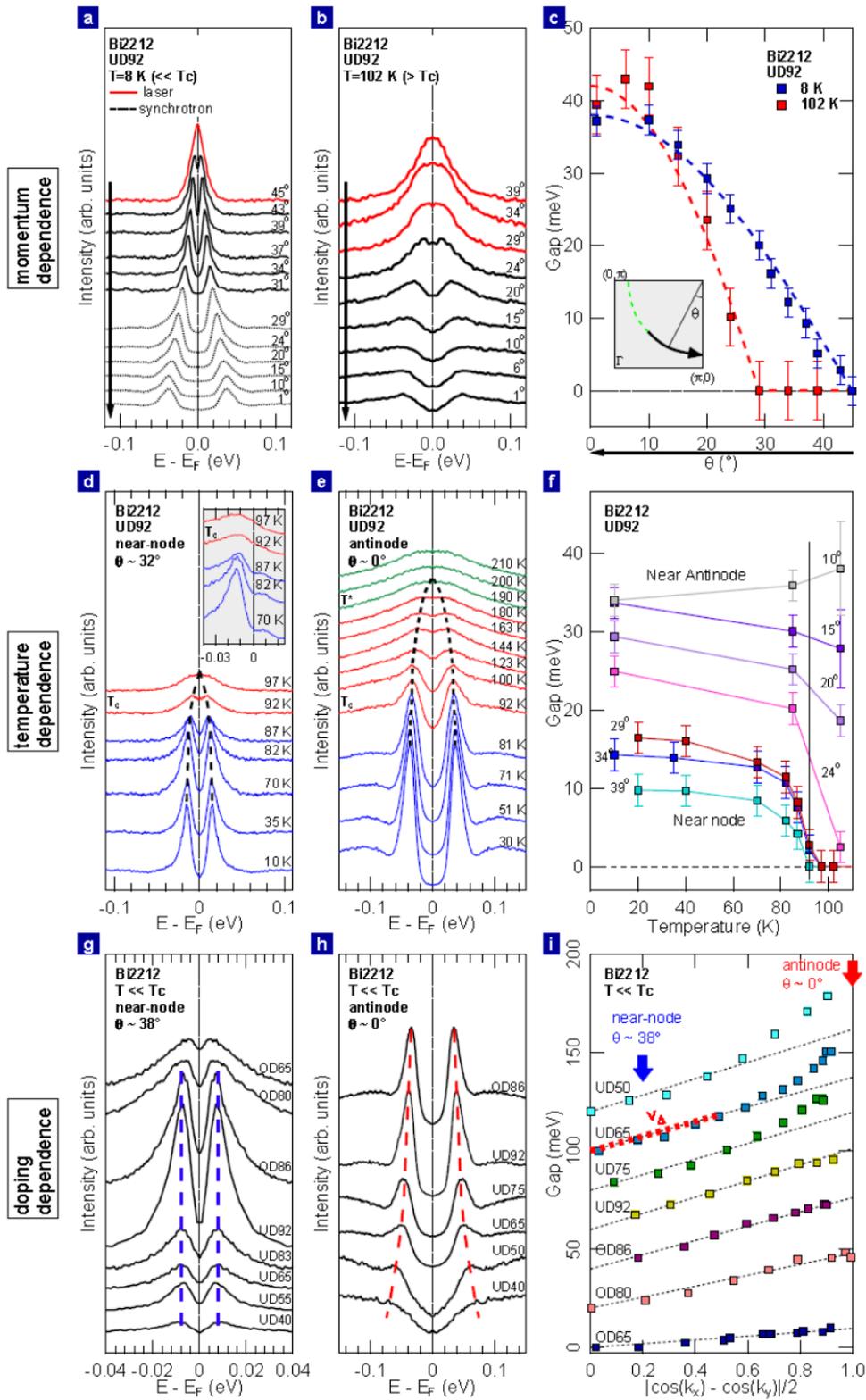

**Fig. 3 Nodal-antinodal dichotomy in Bi2212 a** and **b** Momentum dependence of the symmetrized ARPES spectra for UD92 at $T \ll T_c$ and $T > T_c$, respectively. Red spectra denote ungapped region (Fermi arc). **c** Gap function for UD92K at $T \ll T_c$ and $T > T_c$ plotted as a function of the Fermi angle θ. Gaps near the node close around $T_c$, forming a Fermi arc, whereas the gap magnitude near the antinode does not diminish across $T_c$. **d** and **e** Temperature dependence of the symmetrized spectra near the node and at the antinode, respectively, for UD92. Inset of **d** shows raw spectra that clearly show the upper Bogoliubov peak (right). **f** Temperature dependence of the gap size at various momenta. The gap near the node closes at ∼ $T_c$ following BCS-like behavior whereas the gap around the antinode does not close across $T_c$. The intermediate region shows an intermediate behavior. **g** and **h** Doping dependence of the symmetrized spectra near the node and at the antinode, respectively. Dashed curves in panels **d**, **e**, **g**, and **h** are guides to the eye tracking the peak position, corresponding to the gap size. **i** Doping dependence of the gap function as a function of the *d*-wave form factor $|\cos(k_x)-\cos(k_y)|/2$. Results at each doping level are shown with a vertical offset of 20 meV for clarity. Dashed lines, indicative of a *d*-wave form of the gap, are guides to the eye which emphasize the deviation of the gap function from the *d*-wave form in the underdoped region. The slopes of the dashed lines for UD50, UD65, UD72, and UD92 are fixed. Red dashed line on UD65 indicates $v_\Delta$, and its doping dependence is plotted in Fig. 7c. Part of the data is reproduced from Refs. [14,26,51]. Error bars in uncertainty of determining $E_F$ (±0.5 meV), error from fitting procedure, and an additional 100% margin.

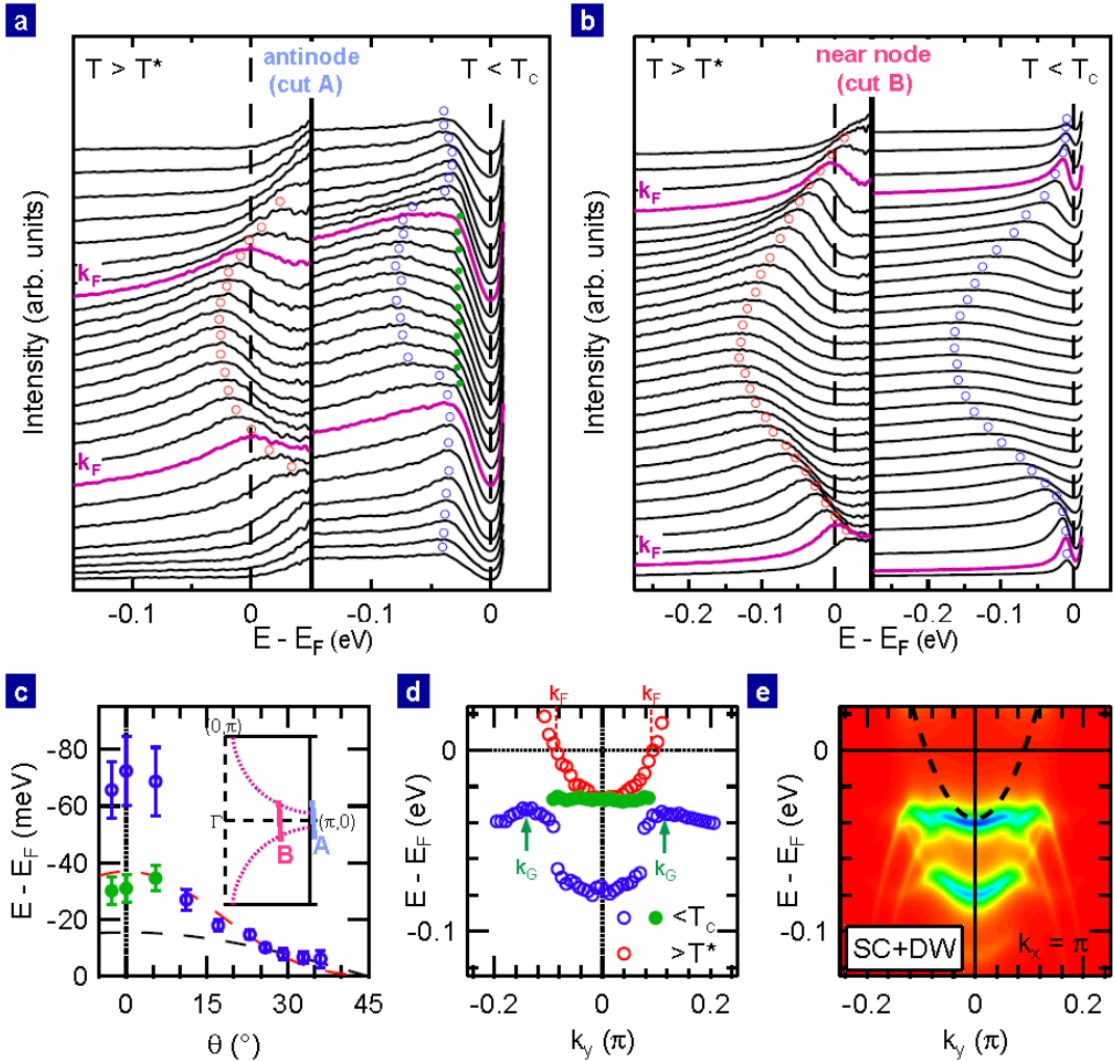

**Fig. 4 Dichotomy between the antinodal and nodal regions in Bi2201 a** and **b** ARPES spectra divided by the Fermi-Dirac function for an antinodal cut **a** and a near-nodal cut **b** at $T > T^*$ and $T < T_c$. **c** Gap function at $T < T_c$. The energy position of the lowest energy feature (green circles in **a**-right) deviates from the simple $d$-wave form near the antinode ($\Delta_{sc}$ = 15.5 meV, black dashed curve), and the intensity maximum is found at higher energy in the antinodal region. Error bars are estimated based on the sharpness of features, based on different EDC analyses.[13] Red dashed curve is from Ref. [38]. Inset (schematic Fermi surface) shows the cuts A and B in panels **a** and **b**,

respectively. **d** Dispersion along the antinodal cut at $T > T^*$ and $T < T_c$. $k_F$ and back-bending momenta ($k_G$) are indicated by red and green arrows, respectively. **e** Renormalized band dispersion by simulations assuming coexistence of *d*-wave superconductivity (order parameter 35 meV) and bond-direction **q**$_1$ = (0.15π, 0) & **q**$_2$ = (0, 0.15π) checkerboard density wave (order parameter 20 meV) along the antinodal cut. Dashed curve is the bare band dispersion from a global tight-binding fit to the experimental dispersions of the intensity maximum at 172 K. All the figures are reproduced from Ref. [13].

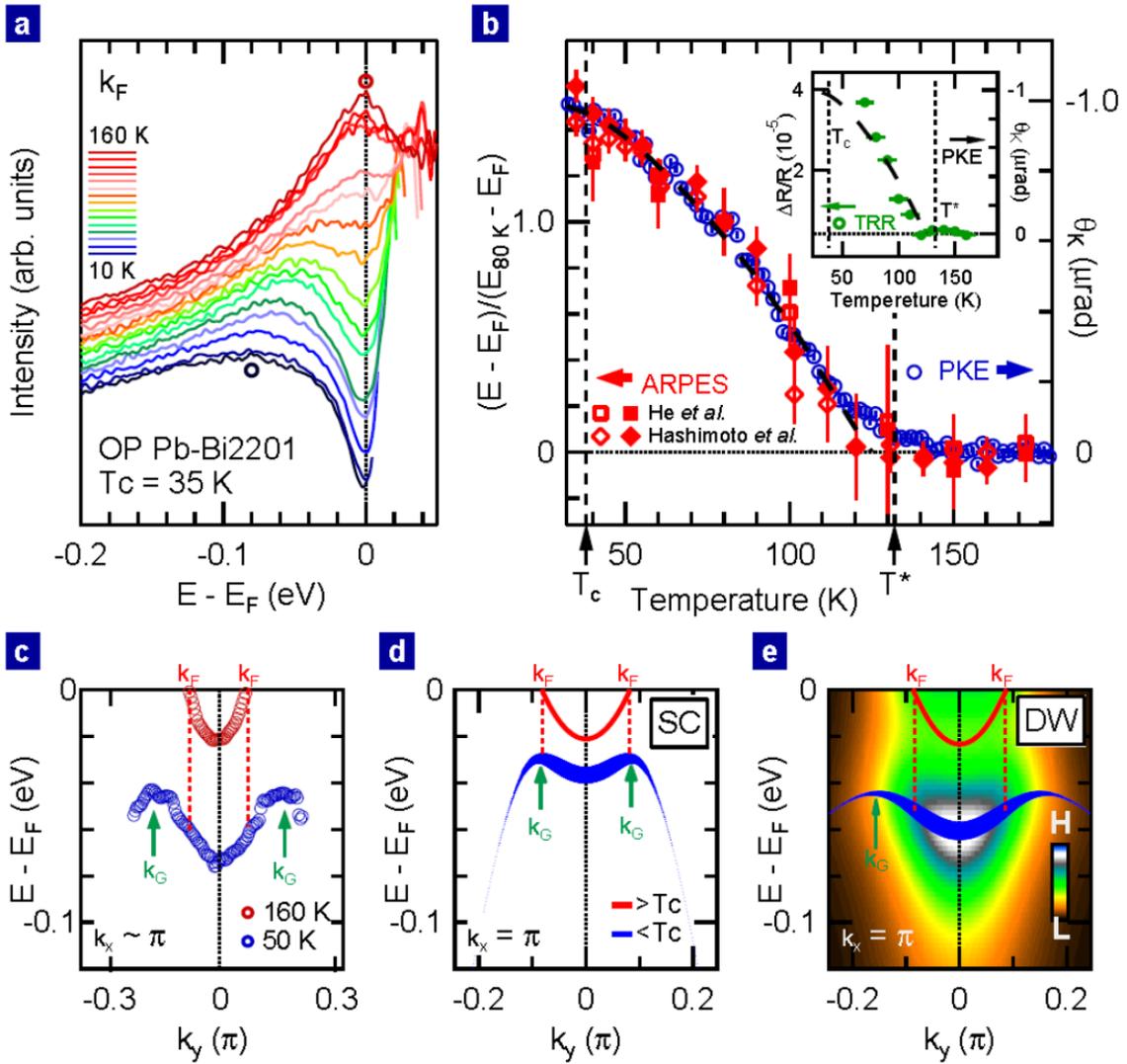

**Fig. 5 Broken-symmetry nature of the pseudogap in Bi2201 a** Temperature dependence of the ARPES spectra at antinodal $k_F$. Blue and red circles indicate the intensity maxima of the spectra at 10 K and 160 K, respectively. **b** Temperature dependence of the energy position of the intensity maximum at $k_F$ given by ARPES, in comparison with the Kerr rotation angle ($\theta_K$) measured by PKE. Inset shows the temperature dependence of the transient reflectivity change measured by TRR (left axis). The dashed black curves (right axis) in the main panel and inset are guides to the eye for the PKE data, showing a mean-field–like critical behavior close to $T^*$. Error bars are estimated based on the sharpness of features, based on different EDC analyses.[13] **c** Dispersions above $T^*$ (red circles) and well below $T^*$ (blue circles). $k_F$ and back-bending momenta are misaligned. **d** and **e** Simulated dispersions for *d*-wave superconductivity (order parameter 30 meV) and antiferromagnetic order (order parameter 60 meV) with a short correlation length (10 unit cell), respectively. The cuts for **c-e** are along $(\pi,-\pi)$–$(\pi, 0)$–$(\pi, \pi)$. $k_F$ and back-bending momenta ($k_G$) in **c-e** are indicated by red and green arrows, respectively. The red (blue) curve is for the true normal (gapped) state. The spectral weight is proportional to the curve thickness. Back-bendings ($k_F$) are indicated by green arrows (red dashed lines). Panels **a** and **c-e** are reproduced from Ref. [12] and **b** from Ref. [13].

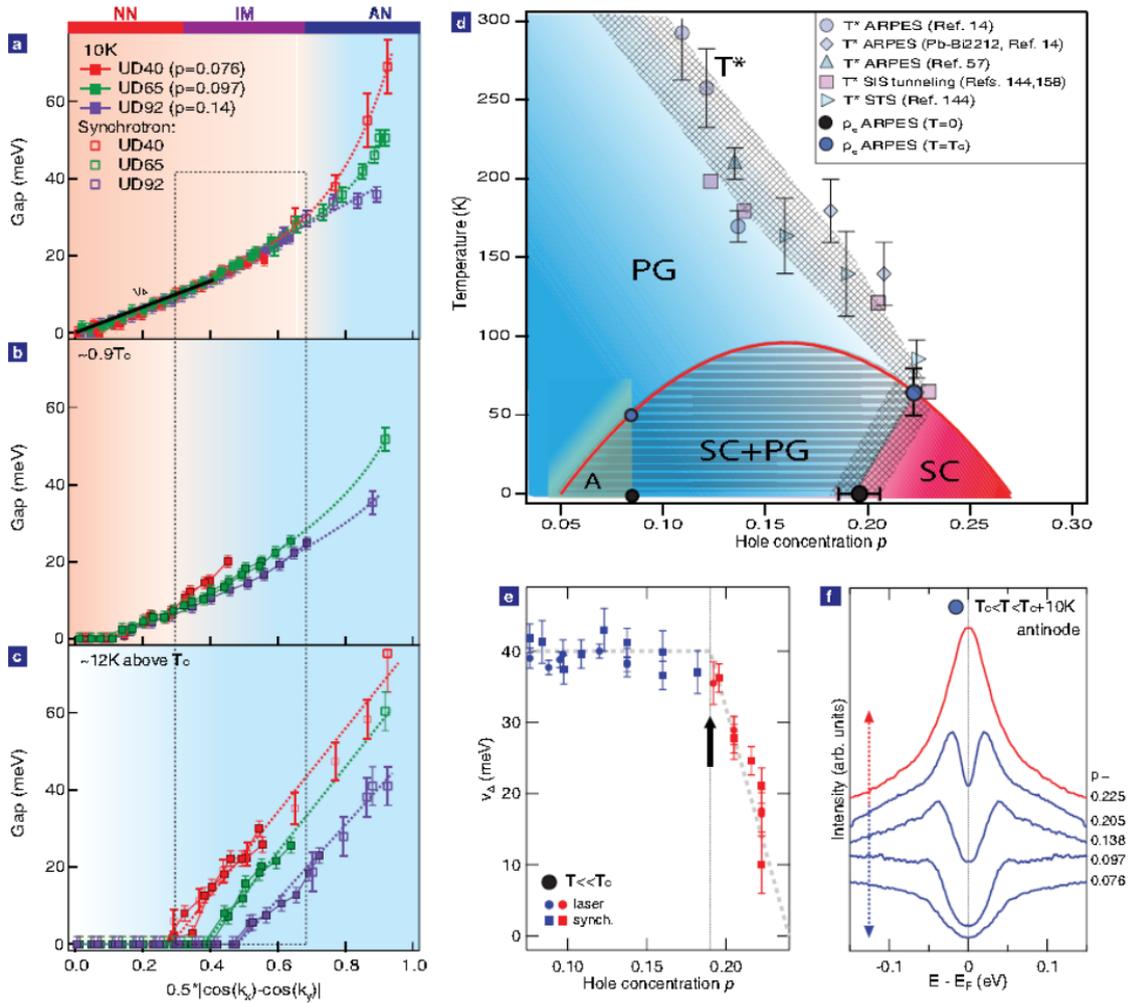

**Fig. 6 Proposed phase diagram of Bi2212 a-c** Phase competition in SC+PG region. Gaps in UD40, UD65, and UD92 at 10K, 0.9 $T_c$, and ~12 K above $T_c$. Synchrotron and laser data are shown with open and filled symbols, respectively. Error bars in laser-ARPES reflect 3σ error in fitting procedure and an additional 100% margin. Error bars in synchrotron data reflect uncertainty of determining $E_F$ (±0.5 meV), error from fitting procedure, and an additional 100% margin. Dashed lines are guides to the eye. Doping-independent or dependent gaps are indicated by pink or blue shading, respectively. Dashed box marks momenta where gaps are doping-dependent in **b** and **c** but doping-independent in **a**. **d** Proposed phase diagram. Superconducting dome is divided into three

phenomenologically distinct regions (ref. [14]): the green-shaded region characterized by a fully gapped Fermi surface, the blue-shaded region where pseudogap order coexists with superconductivity (SC+PG), and the red-region region where the pseudogap order is absent below $T_c$ (SC). $T^*$ is determined from ARPES measurements at antinode[14,57], STS[146], and superconductor-insulator-superconductor tunneling.[144,158] $T^*$ which is higher than measurement temperature is estimated from an extrapolation of the antinodal gap size. Error bars in $T^*$ is 3σ in linear fit. For $T^*$ which is accessible by ARPES, error bars are temperature interval between data points[14]. **e** Doping dependence of the symmetrized antinodal spectra slightly above $T_c$, indicating the existence of the pseudogap at least up to $p \sim 0.22$. **f** Doping dependence of $v_\Delta$ (see panel **a**) at $T \ll T_c$. $v_\Delta$ shows an abrupt change at $p \sim 0.19$ (indicated by an arrow), which is interpreted as the $T = 0$ endpoint of the pseudogap order. $v_\Delta$ is from a fit over the linear portion of the gap function, as shown by a solid line in **a**. Error bar in $v_\Delta$ is 3σ confidence interval for slope. Figures are adapted from Ref. [14] with some data points from more recent experiments added.